\documentclass[twocolumn]{aastex62}

\graphicspath{{./}{figures/}}

\accepted{to ApJL}

\shorttitle{Sample article}
\shortauthors{Mezcua, M. \& Dom\'inguez S\'anchez, H.}

\begin{document}

\title{Hidden AGN in dwarf galaxies revealed by MaNGA: light echoes, off-nuclear wanderers, and a new broad-line AGN}

\correspondingauthor{Mar Mezcua}
\email{marmezcua.astro@gmail.com}

\author[0000-0003-4440-259X]{Mar Mezcua}
\affiliation{Institute of Space Sciences (ICE, CSIC), Campus UAB, Carrer de Magrans, 08193 Barcelona, Spain}
\affiliation{Institut d'Estudis Espacials de Catalunya (IEEC), Carrer Gran Capit\`{a}, 08034 Barcelona, Spain}

\author{Helena Dom\'inguez S\'anchez}
\affiliation{Institute of Space Sciences (ICE, CSIC), Campus UAB, Carrer de Magrans, 08193 Barcelona, Spain}



\begin{abstract}
Active galactic nuclei (AGN) in dwarf galaxies could possibly host the relics of those early Universe seed black holes that did not grow into supermassive black holes. Using MaNGA integral field unit (IFU) spectroscopy we have found a sample of 37 dwarf galaxies that show AGN ionisation signatures in spatially-resolved emission line diagnostic diagrams. The AGN signatures are largely missed by integrated emission line diagnostics for 23 of them. The bolometric luminosity of these 23 new AGN candidates is $\lesssim 10^{40}$ erg s$^{-1}$, fainter than that of single-fiber SDSS AGN, X-ray AGN, and radio AGN in dwarf galaxies, which stands IFU spectroscopy as a powerful tool for identifying hidden and faint AGN in dwarf galaxies. 
The AGN emission is in most cases offset from the optical center of the dwarf galaxy and shows a symmetric morphology, which indicates that either the AGN are off-nuclear, that the central emission of the galaxy is dominated by star formation, or that the AGN are turned-off and we are observing a past ionisation burst. One of the new AGN shows a broad H$\alpha$ emission line component, from which we derive a black hole mass in the realm of intermediate-mass black holes. This constitutes the first hidden type 1 AGN discovered in a dwarf galaxy based on IFU spectroscopy.
The finding of this sample of hidden and faint AGN has important implications for population studies of AGN in dwarf galaxies and for seed black hole formation models.
\end{abstract}

\keywords{Active galactic nuclei -- Dwarf galaxies}


\section{Introduction} 
\label{intro}
The finding of hundreds of active galactic nuclei (AGN) in low-mass, dwarf galaxies (stellar mass $M_\mathrm{*} \leq 3 \times 10^{9}$ M$_{\odot}$) has revolutionized black hole and galaxy formation models. These AGN are often found to be powered by low-mass or intermediate-mass black holes (IMBHs; $100 < M_\mathrm{BH} \lesssim 10^{6}$ M$_{\odot}$), which could be the relics of the early Universe seed black holes from which supermassive black holes ($M_\mathrm{BH} > 10^{6}$ M$_{\odot}$) grow (see \citealt{2017IJMPD..2630021M}; \citealt{2019arXiv191109678G} for recent reviews). If these seeds have not significantly grown by mergers or AGN feedback through cosmic time (\citealt{2019NatAs...3....6M,2020arXiv200411911M}), characterising the local population of IMBHs in dwarf galaxies, in particular the black hole occupation fraction and the low-mass end of the scaling relations, can tell us how seed black holes predominantly formed at $z>$10 (e.g. \citealt{2010MNRAS.408.1139V}; \citealt{2018MNRAS.481.3278R}). 

A myriad of studies have hence focused on identifying AGN in dwarf galaxies. The first searches were based on optical spectroscopic surveys such as the Sloan Digital Sky Survey (SDSS\footnote{\url{https://www.sdss.org}}) and the use of narrow emission-line diagnostic diagrams (BPT diagram; \citealt{1981PASP...93....5B}; \citealt{2001ApJ...556..121K,2006MNRAS.372..961K}; \citealt{2003MNRAS.346.1055K}) to distinguish between gas ionisation by AGN o by stars, accompanied by the detection of broad Balmer emission lines from which a black hole mass measurement could be obtained (e.g. \citealt{2013ApJ...775..116R}; \citealt{2014AJ....148..136M}). Such optical studies are however biased towards central luminous type-1 AGN with high accretion rates and at low redshifts (z $<$ 0.3). This can be circumvented by X-ray and radio searches, which are not only able to detect AGN in dwarf galaxies at the pinnacle of cosmic star formation history (\citealt{2016ApJ...817...20M,2018MNRAS.478.2576M,2019MNRAS.488..685M}), but offer the most reliable measurements of the AGN fraction in dwarf galaxies (e.g. \citealt{2018MNRAS.478.2576M}; \citealt{2020MNRAS.492.2268B}). 

In addition to the biases of optical searches, single-fibre BPT diagnostics can fail at identifying AGN in dwarf galaxies with active star formation (e.g. \citealt{2019ApJ...870L...2C}; \citealt{2020MNRAS.492.2268B}) or with a strong host galaxy light that dilutes the AGN signatures (e.g. \citealt{2002ApJ...579L..71M}). This can be circumvented by high-resolution optical spectroscopy (e.g. \citealt{2019ApJ...884..180D}) or optical variability studies (e.g. \citealt{2018ApJ...868..152B,2020ApJ...896...10B}), which are able to reveal AGN signatures even in quiescent and star-formation-dominated dwarf galaxies.

The advent of integral-field unit (IFU) spectroscopic surveys such as SDSS/MaNGA (Mapping Nearby Galaxies at APO; \citealt{2015ApJ...798....7B}) has opened a new window in the identification of `hidden' AGN (e.g. \citealt{2018MNRAS.474.1499W}). Such sources are missed by single-fibre BPT diagnostics if other mechanisms rather than the AGN dominate the ionisation at the galaxy center, if the AGN is displaced from the central region due to a galaxy merger (e.g. \citealt{2014ApJ...789..112C}; \citealt{2018ApJ...869..154B}), or if the AGN has recently switched off and its ionisation signatures are only observable as light echoes at large distances from the optical center (e.g. \citealt{2015AJ....149..155K}). Yet, most AGN searches using IFU data are biased towards central AGN or do not focus on dwarf galaxies (\citealt{2017MNRAS.472.4382R}; \citealt{2018RMxAA..54..217S}; \citealt{2018MNRAS.474.1499W}). \cite{2018MNRAS.476..979P} used the MaNGA survey to investigate AGN in dwarf galaxies, but their sample included only 69 quenched systems. Hence, no systematic studies of AGN in dwarf galaxies has been performed so far using IFU data. 

In this Letter we present the largest dedicated study of AGN in dwarf galaxies using MaNGA data of 4718 sources. This allows us to identify AGN candidates in 37 dwarf galaxies, in 23 of which no clear AGN signatures are revealed in the single-spectrum BPT, including the discovery of a new broad-line low-mass AGN. The sample and analysis are presented in Sect.~\ref{sample}, while the results obtained are discussed in Sect.~\ref{results}. Final conclusions are provided in Sect.~\ref{conclusions}.

\section{Sample and analysis} 
\label{sample}
The sample of dwarf galaxies studied in this paper is drawn from the SDSS/MaNGA survey Data Release\footnote{\url{http://www.sdss.org/surveys/manga}} 15, which includes 4718 galaxies. The MaNGA IFU is centered on the optical center of the galaxy and has a size that typically matches the angular size of the galaxy, so that most targets are covered spectroscopically to at least 1.5 effective radii ($R_\mathrm{eff}$). 
The final datacubes provided by the MaNGA data-analysis pipeline\footnote{\url{https://www.sdss.org/dr15/manga/manga-analysis-pipeline/}} have square spaxels of 0.5 arcsec size, and the median spatial resolution is of 2.5 arcsec Full Width Half Maximum (FWHM). In this paper we use the emission line measurements (performed after subtraction of the stellar continuum and corrected for absorption) of the HYB10-GAU-MILESHC maps, in which the spaxels are binned to a signal-to-noise ratio (SNR) $\sim$10 but the kinematic properties are extracted from the individual spaxels.

We also take from the MaNGA data-analysis pipeline the star-formation rate (SFR) within the IFU field-of-view based on the H$\alpha$ flux and the flux-weighted mean stellar velocity dispersion ($\sigma$) of all spaxels within 1 $R_\mathrm{eff}$. 

The $R_\mathrm{eff}$, redshift ($z$) and $M_{*}$ are taken from the NASA-Sloan Atlas (NSA catalog) version v1\_0\_1, as used by MaNGA for its targeting\footnote{\url{https://www.sdss.org/dr16/manga/manga-target-selection/nsa/}}, which incorporates an expanded redshift range and an improved elliptical Petrosian aperture photometry compared to the previous NSA catalog v0\_1\_2. The SFR and $M_{*}$ are provided in units of h$^{-2}$ M$_{\odot}$ yr$^{-1}$ and h$^{-2}$ M$_{\odot}$, respectively, and we assume h=0.73. 

To draw our sample we first select those sources for which the Petrosian and Sersic NSA stellar masses are consistent within 0.5 dex (e.g. \citealt{2020arXiv200506452S}). The target dwarf galaxies are then selected as having $M_\mathrm{*} < 3 \times 10^{9}$ M$_{\odot}$, as commonly done in AGN studies of dwarf galaxies (e.g. \citealt{2013ApJ...775..116R,2020ApJ...888...36R}; \citealt{2018MNRAS.478.2576M,2019MNRAS.488..685M}). No magnitude cut is applied to avoid any biases towards galaxy type. This yields an initial sample of 1609 dwarf galaxies.

\begin{figure*}
\centering
\includegraphics[width=\textwidth]{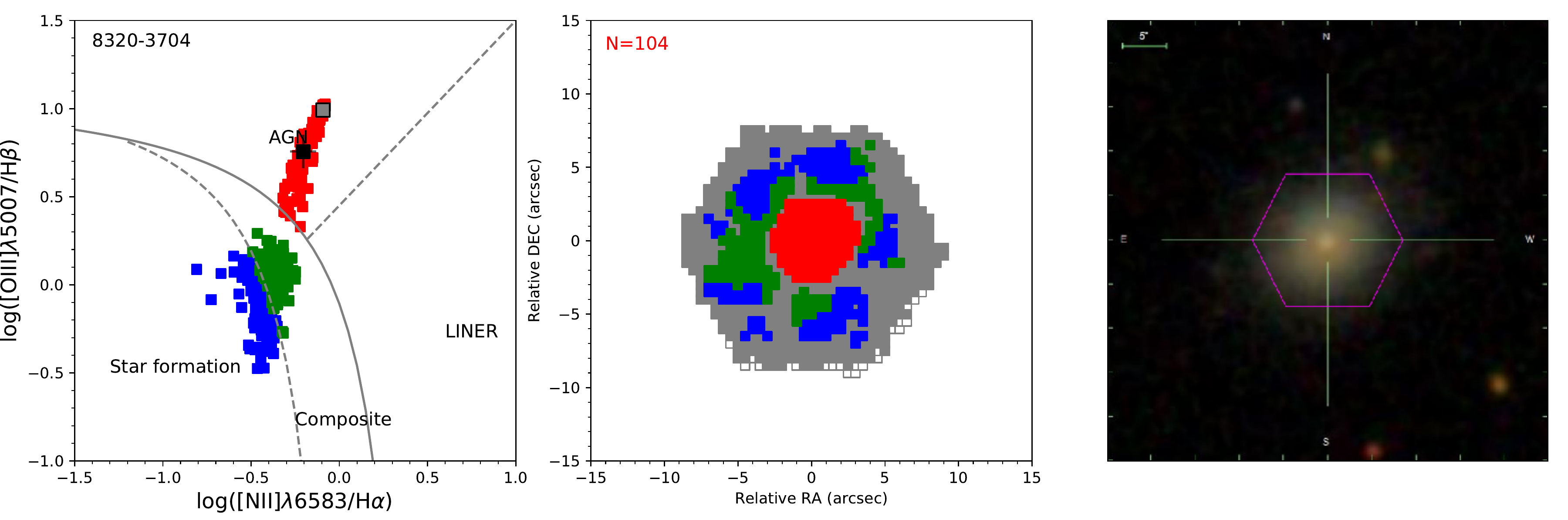}
\includegraphics[width=\textwidth]{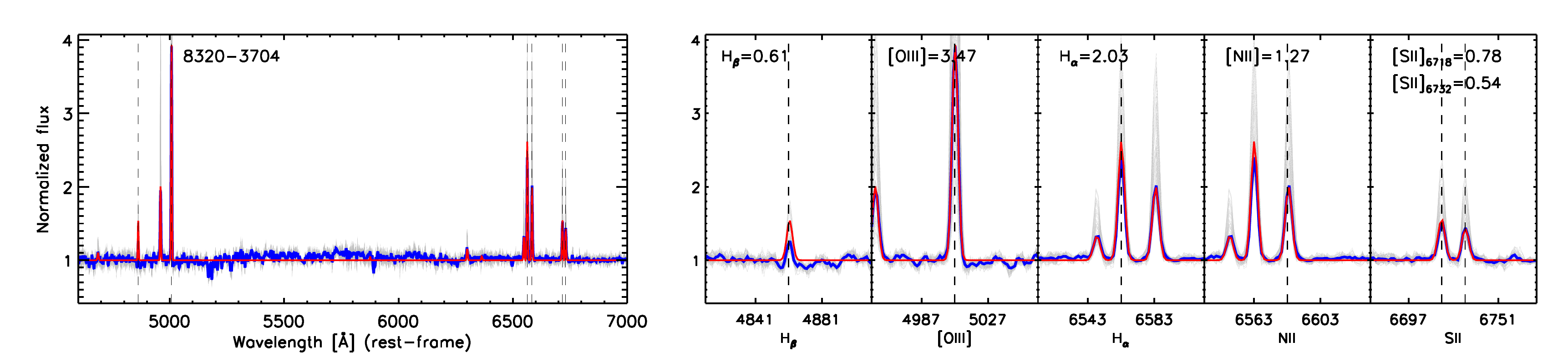}
\includegraphics[width=\textwidth]{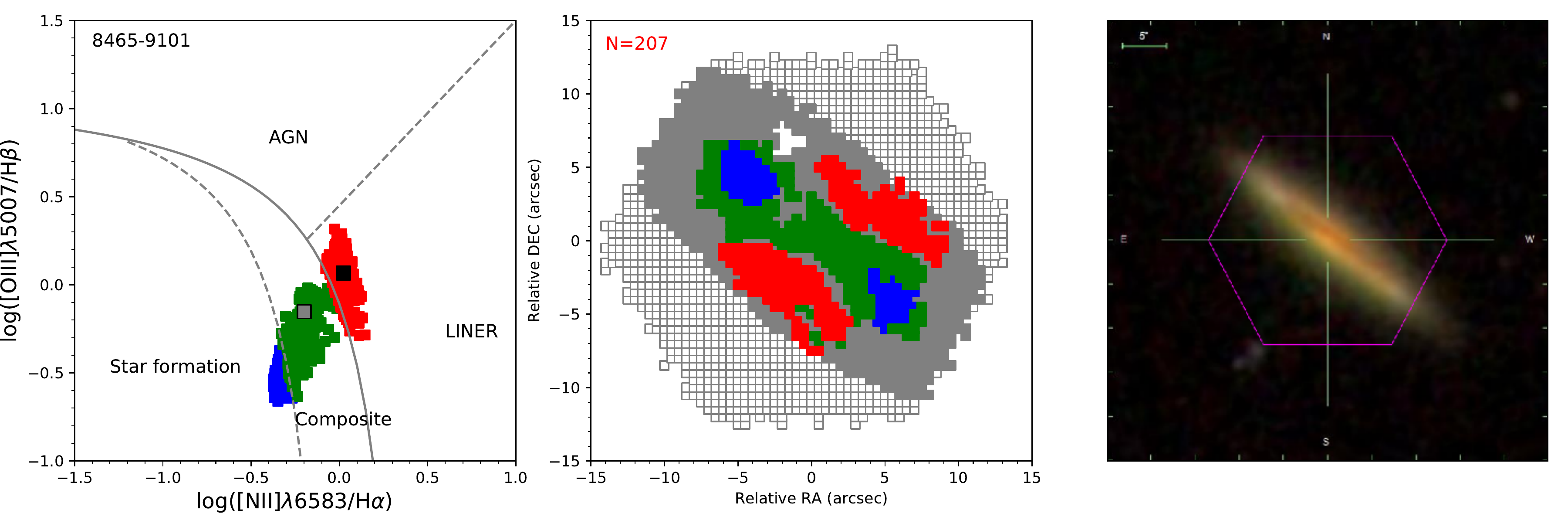}
\includegraphics[width=\textwidth]{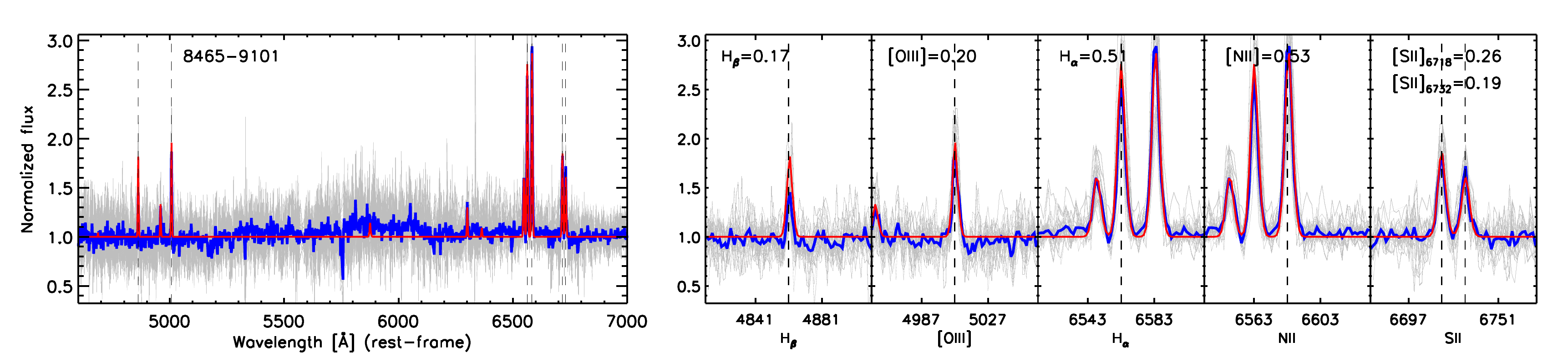}
\caption{MaNGA analysis for four of the AGN dwarf galaxy candidates. For each source: \textbf{Top left}: Location of each MaNGA spaxel on the [NII]-BPT used to distinguish between ionisation by AGN/LINER (red spaxels), star-formation (blue spaxels), or composite (green spaxels). The black square marks the median BPT location of those spaxels classified as AGN/LINER, the grey square the SDSS (single-fiber) BPT location; \textbf{Top center}: Spatial distribution of the BPT-classified spaxels (color-coded as in the left panel). Empty squares mark the IFU coverage, grey squares those spaxels with continuum SNR $>$ 1. The N shows the number of AGN/LINER spaxels used in the analysis and stacking; \textbf{Top right}: SDSS composite image. The pink hexagon shows the IFU coverage; \textbf{Bottom left}: Stacked spectrum (in blue) of all the galaxy spaxels (in grey) which fall in the AGN/LINER region of the BPT diagram. The emission line component is shown in red; \textbf{Bottom right}: Zoom-in of the stack in the region of the emission lines used in the BPT. The median emission line flux of the AGN/LINER spaxels is shown in units of 10$^{-17}$ erg s$^{-1}$ cm$^{-2}$.}
\label{fig1}
\end{figure*}

\begin{figure*}
\centering
\includegraphics[width=\textwidth]{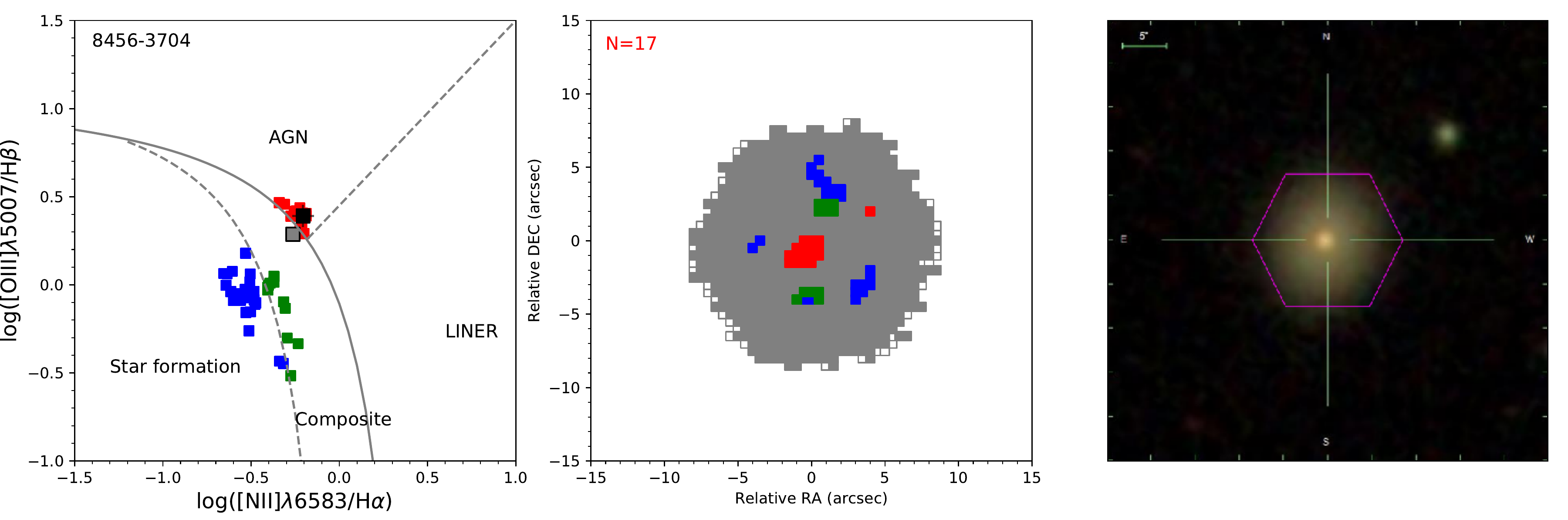}
\includegraphics[width=\textwidth]{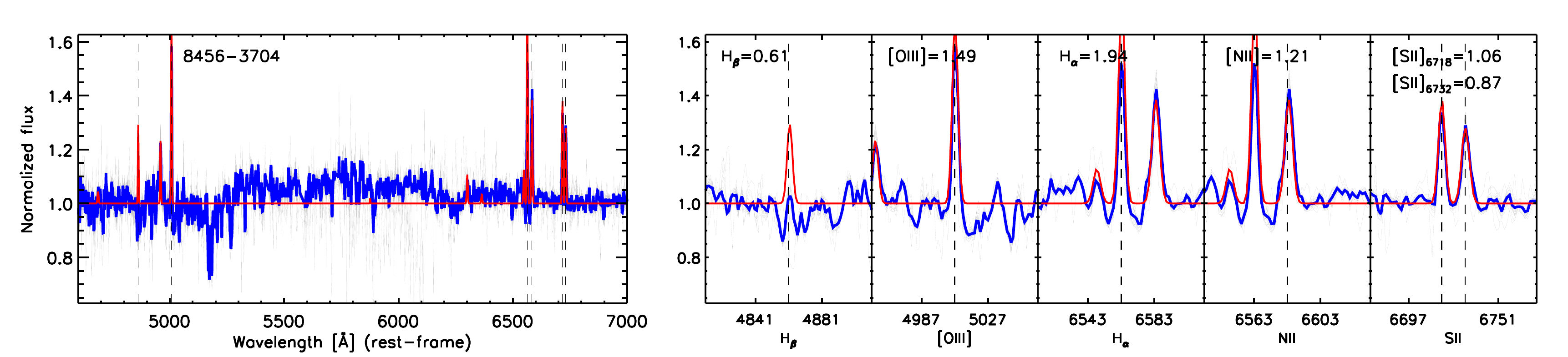}
\includegraphics[width=\textwidth]{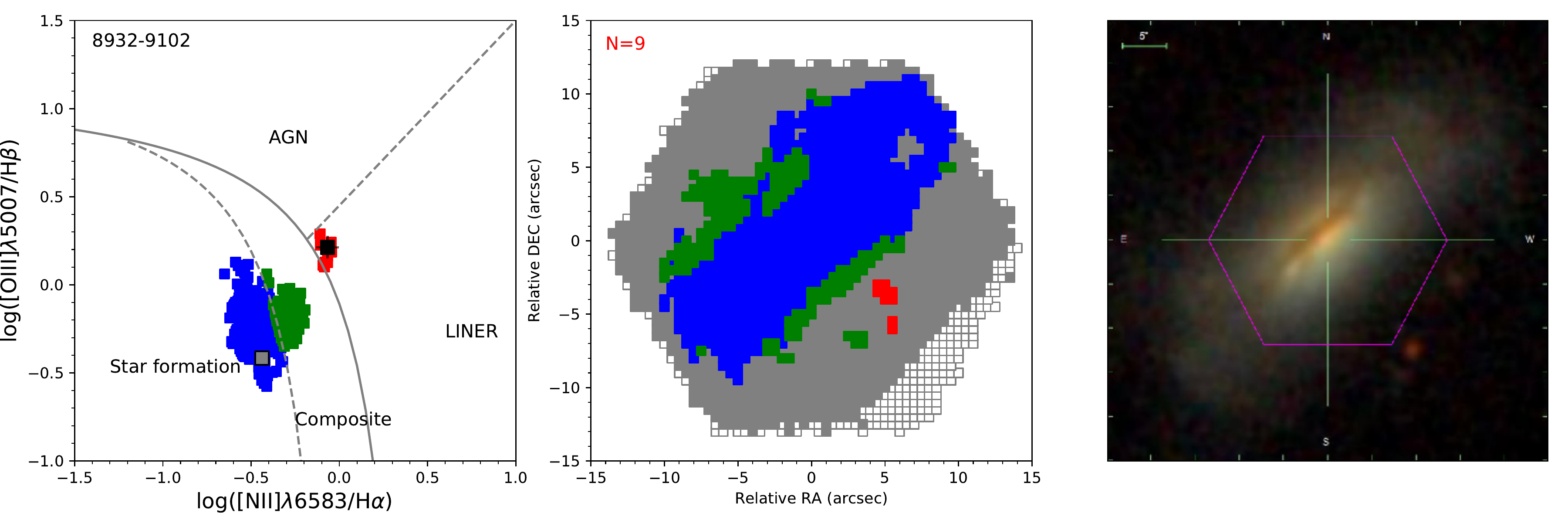}
\includegraphics[width=\textwidth]{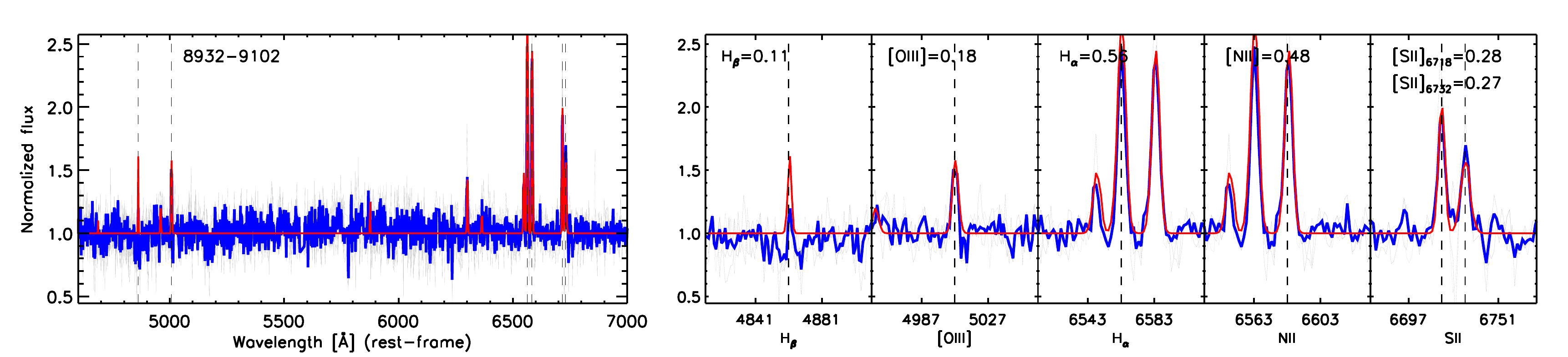}
Figure~\ref{fig1} (cont.)
\end{figure*}

To identify which is the source of ionisation of each spaxel we use the lines of \cite{2001ApJ...556..121K,2006MNRAS.372..961K} and \cite{2003MNRAS.346.1055K} to distinguish between AGN, star formation, or composite (combination of AGN and star formation ionisation) in the [OIII]/H$_{\beta}$ versus [NII]/H$_{\alpha}$ ([NII]-BPT) and [OIII]/H$_{\beta}$ versus [SII]/H$_{\alpha}$ ([SII]-BPT) emission line diagnostics. We also use the line from \cite{2007MNRAS.382.1415S} for a LINER\footnote{Low Ionization Emission Line Region (LINER)} division in the [NII]-BPT. We only consider those dwarf galaxies for which the SNR of the lines used in the [NII]-BPT and [SII]-BPTdiagrams (H$_{\alpha}$, H$_{\beta}$, [NII]$\lambda$6583, [SII]$\lambda$6718, [OIII]$\lambda$5007) is $\geq$3. For these we first plot the [NII]-BPT and [SII]-BPT and select those galaxies for which at least 5 spaxels are located in the AGN/LINER photoionisation region of both these two diagram (see Fig.~\ref{fig1}, top left panel). Then, for each galaxy, the individual spectra of all the spaxels identified as AGN or LINER according to the above criteria are stacked together in order to increase the signal. The stacking procedure is explained in detail in Appendix \ref{stacking}. Those sources whose stacked spectrum is very noisy and/or for which the emission lines of interest are hardly identifiable or have inconsistent measurements are removed from the sample. We also exclude those galaxies for which the spaxels classified as AGN/LINER are spread at the edges of the SNR $>$ 1 continuum map (grey area in Fig.~\ref{fig1}, top middle panel). This yields a sample of 102 AGN dwarf galaxy candidates.

\begin{figure}
\centering
\includegraphics[width=0.49\textwidth]{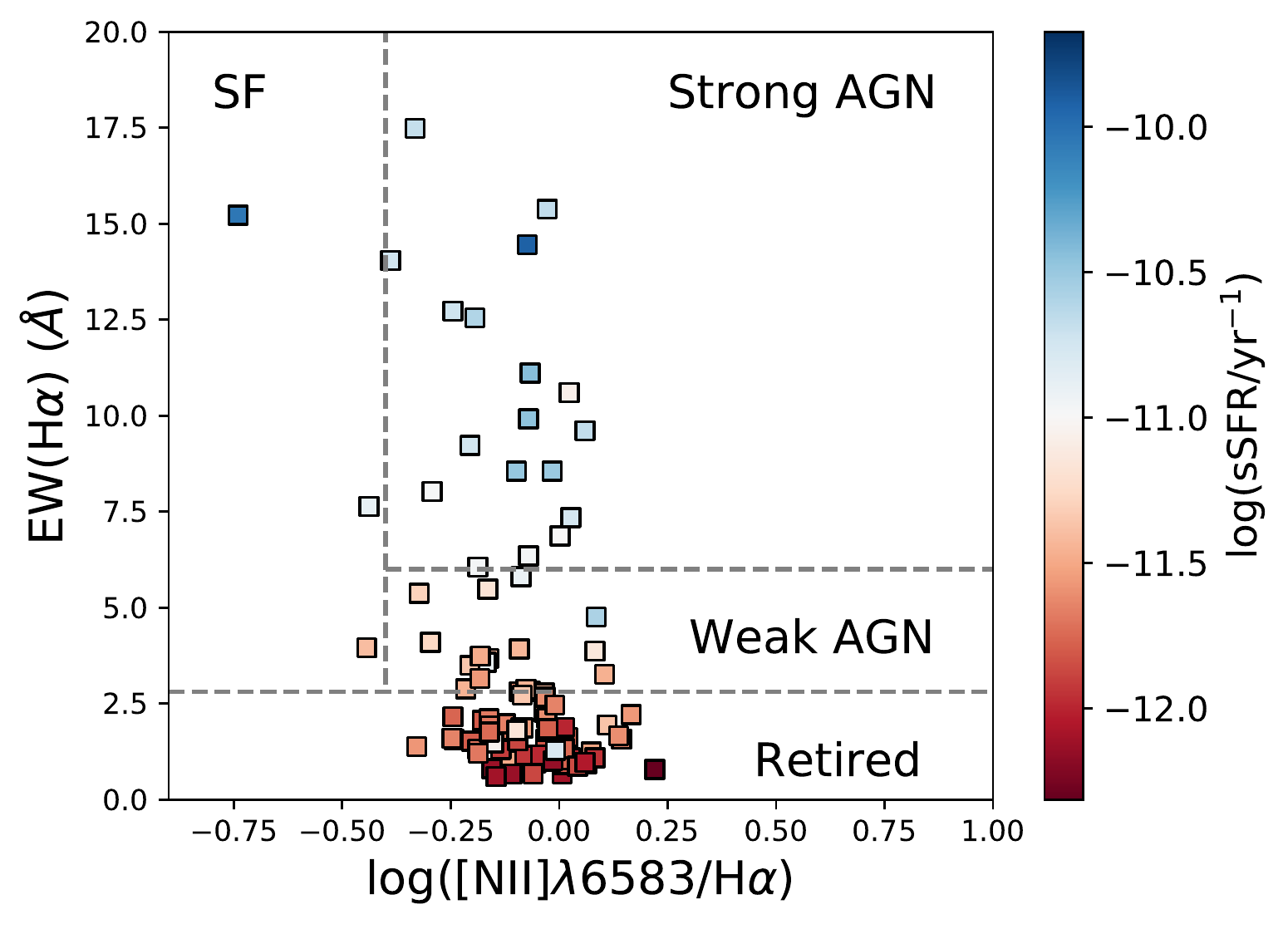}
\caption{WHAN diagram for the initial 102 dwarf galaxies classified as AGN/LINER according to the  [NII]-BPT and [SII]-BPT diagrams. The final sample of AGN dwarf galaxy candidates is formed by the 37 sources classified as strong or weak AGN in the WHAN diagram. The color bar denotes the median sSFR of the AGN/LINER spaxels.}
\label{WHAN}
\end{figure}

We derive a median BPT location of the 102 AGN candidates by taking the median of the [NII]-BPT line fluxes of the same AGN/LINER spaxels used to obtain the stacked spectrum (black square in Fig.~\ref{fig1}, top left panel). Based on this median BPT location, 47 of the AGN candidates are classified as LINERs. The LINER emission can not only originate from AGN but also be produced on galactic scales by hot old stars, which are expected to have an H$\alpha$ equivalent width (EW) typically below 3\AA. A distinction between the two processes can be made using the WHAN diagram (\citealt{2010MNRAS.403.1036C}), which suggests that sources with EW(H$\alpha$) $>$ 3\AA\ and flux ratio log [NII]/H$\alpha >$ -0.4 can be classified as true AGN. A further sub-classification can be performed among strong AGN (EW(H$\alpha$) $>$ 6\AA) and weak AGN (3\AA\ $<$ EW(H$\alpha$) $<$ 6\AA ; see Fig.~\ref{WHAN}). Based on the WHAN diagram of the median value of the AGN/LINER spaxels and relaxing the EW(H$\alpha$) cut to $>$ 2.8\AA\ in order not to exclude bona-fide AGN (see Sect.~\ref{results}), we find that the AGN/LINER spaxels of 37 out of the 102 AGN candidates are consistent with being dominated by AGN ionisation. These 37 sources constitute our final sample of AGN dwarf galaxy candidates.

\section{Results and discussion} 
\label{results}
Of the 37 AGN candidates identified above using MaNGA data, 36 have also single-fibre (3 arcsec aperture) SDSS spectra available. For one of the sources (9881-1901) the H$\alpha$ region is missing, so we consider only 35 SDSS spectra. Two of them (8446-1901 and 9000-1901) are known broad-line AGN (e.g. \citealt{2015ApJ...813...82R}, from which we take the narrow emission line fluxes to obtain the SDSS BPT classification), while for the remaining 33 sources we obtain a SDSS BPT classification based on the emission-line measurements of the SDSS DR12 spectra provided by the Portsmouth Group$\footnote{\url{https://www.sdss.org/dr12/spectro/galaxy_portsmouth/}}$ catalog. We classify the sources as Quiescent if the SNR of any of the SDSS H$_{\alpha}$, H$_{\beta}$, [NII]$\lambda$6584, or [OIII]$\lambda$5007 emission lines used in the BPT is $<$3 (e.g. \citealt{2020arXiv200506452S}). If the SNR $\geq$ 3, the sources are classified as either Star-forming, AGN, or Composite following the same classification as for the MaNGA BPT (see Sect.~\ref{sample}). We find that twelve out of the 33 sources are classified as AGN according to the SDSS (single-fiber) BPT, in agreement with the MaNGA BPT classification. However, two of these AGN have a median MaNGA EW(H$\alpha\sim$2.8\AA) and would be missed by the WHAN diagram classification. To ensure we do not exclude any bona fide AGN we thus relax the EW(H$\alpha$) cut to $>$ 2.8\AA. The remaining 21 sources out of the 33 are classified as either Quiescent, Star-forming or Composite.  Using IFU MaNGA data we are thus able to identify 23 new AGN in dwarf galaxies whose AGN signatures are hidden (or partly hidden in the case of Composite objects) or not available in the BPT of single-fibre SDSS spectra. The use of integrated spectra to search for AGN can therefore fail in low-mass galaxies, biasing the demographics of dwarf galaxies hosting an AGN (see Sect.\ref{demographics}; e.g. \citealt{2013ApJ...775..116R}; \citealt{2019ApJ...870L...2C}; \citealt{2020MNRAS.492.2268B}).

\begin{table*}
\footnotesize{}
\caption{Properties of the 37 AGN candidates in dwarf galaxies.}
\label{table1}
\begin{tabular}{lcccccccccc}
\hline
\hline 
MaNGA & RA & DEC & $z$ &  log $M_\mathrm{*}$ & log $SFR$ & TType & $R_\mathrm{eff}$ & log $L_\mathrm{bol}$ & MaNGA  & SDSS \\
plateifu & (J2000) & (J2000) &  & (M$_{\odot}$) &  (M$_{\odot}$ yr$^{-1}$) & & (kpc) & (erg s$^{-1}$) & BPT & BPT\\
(1) & (2)   & (3)   & (4)   & (5)   & (6)   & (7)   &  (8)  & (9)   & (10) & (11)\\
\hline
  7958-9101$^{*}$  &  258.495841  &  33.607137  &  0.039  &      9.36  &    -1.32  &    0.9  &       3.8  &  40.8  &       AGN  &           AGN  \\
        7960-3703  &  259.183520  &  32.806265  &  0.041  &      9.26  &    -1.33  &   -1.3  &       2.5  &  39.9  &       AGN  &  Star-Forming  \\
        7990-3702  &  262.100040  &  57.002099  &  0.030  &      9.16  &    -1.84  &    1.6  &       2.6  &  39.4  &     LINER  &     Quiescent  \\
        8083-3704  &   51.284736  &  -0.544948  &  0.038  &      9.29  &    -2.17  &    0.4  &       2.2  &  40.0  &     LINER  &     Composite  \\
        8149-1901  &  120.221693  &  28.130394  &  0.017  &      9.26  &    -2.01  &   -2.1  &       1.3  &  40.5  &       AGN  &     Composite  \\
 8255-12704$^{*}$  &  165.117740  &  44.260967  &  0.025  &      9.46  &    -1.21  &    3.9  &       4.4  &  39.3  &     LINER  &  Star-Forming  \\
        8311-3701  &  203.777234  &  22.661833  &  0.032  &      9.48  &    -1.98  &   -1.7  &       1.6  &  40.2  &       AGN  &           AGN  \\
        8320-3702  &  205.289300  &  23.277928  &  0.027  &      9.30  &    -1.17  &    3.5  &       2.1  &  39.3  &     LINER  &  Star-Forming  \\
        8320-3704  &  206.612456  &  22.076742  &  0.028  &      9.01  &    -1.73  &    1.0  &       2.6  &  40.7  &       AGN  &           AGN  \\
  8442-1901$^{*}$  &  199.675905  &  32.918662  &  0.036  &      9.17  &    -1.02  &    0.3  &       1.2  &  40.2  &       AGN  &     Composite  \\
        8446-1901  &  205.753337  &  36.165656  &  0.024  &      9.18  &    -1.40  &   -2.1  &       0.9  &  40.0  &     LINER  &           AGN  \\
       8448-12704  &  166.739672  &  22.835797  &  0.023  &      9.32  &    -1.44  &    4.6  &       4.3  &  39.2  &     LINER  &     Composite  \\
        8452-1901  &  155.885556  &  46.057755  &  0.026  &      9.36  &    -2.07  &    2.1  &       2.0  &  40.5  &       AGN  &           AGN  \\
        8453-1901  &  153.365546  &  47.516236  &  0.025  &      9.40  &    -1.50  &    1.0  &       1.7  &  39.6  &       AGN  &     Composite  \\
       8456-12705  &  151.284662  &  44.514047  &  0.026  &      9.42  &    -1.06  &    5.5  &       7.1  &  39.3  &     LINER  &  Star-Forming  \\
        8456-3704  &  150.228406  &  44.764793  &  0.028  &      9.25  &    -2.11  &    0.4  &       2.1  &  40.4  &       AGN  &     Composite  \\
  8465-9101$^{*}$  &  197.580687  &  47.124056  &  0.024  &      9.46  &    -1.61  &    3.8  &       3.7  &  39.4  &     LINER  &     Composite  \\
        8466-1901  &  168.480327  &  45.488367  &  0.029  &      9.38  &    -2.14  &    1.3  &       1.5  &  40.0  &     LINER  &           AGN  \\
        8547-6104  &  219.546188  &  53.462575  &  0.038  &      9.22  &    -1.75  &    0.7  &       2.7  &  39.9  &       AGN  &     Composite  \\
       8549-12704  &  242.978645  &  46.127727  &  0.020  &      9.20  &    -1.47  &    3.4  &       3.4  &  38.9  &     LINER  &             -  \\
        8588-1901  &  249.717086  &  40.199348  &  0.036  &      9.18  &    -1.32  &   -0.8  &       1.3  &  39.6  &     LINER  &  Star-Forming  \\
        8600-6102  &  244.658478  &  41.136756  &  0.038  &      9.44  &    -1.70  &    3.0  &       3.4  &  39.6  &     LINER  &     Quiescent  \\
        8655-6103  &  355.825111  &   0.442475  &  0.037  &      9.28  &    -2.28  &    1.7  &       2.6  &  40.5  &       AGN  &           AGN  \\
        8715-6103  &  118.872451  &  52.010499  &  0.022  &      9.42  &    -2.07  &    3.8  &       2.9  &  39.8  &       AGN  &     Composite  \\
        8720-1901  &  121.147928  &  50.708556  &  0.023  &      9.39  &    -1.90  &   -2.0  &       1.7  &  40.6  &       AGN  &           AGN  \\
  8932-9102$^{*}$  &  196.651709  &  27.872946  &  0.021  &      9.44  &    -0.99  &    4.1  &       3.5  &  39.2  &     LINER  &  Star-Forming  \\
        8937-1901  &  116.997855  &  29.190689  &  0.027  &      8.96  &    -2.03  &    1.0  &       1.6  &  40.2  &       AGN  &     Composite  \\
  8982-3703$^{*}$  &  203.190094  &  26.580376  &  0.047  &      8.88  &    -0.79  &    1.1  &       1.2  &  41.4  &       AGN  &           AGN  \\
       8990-12705  &  173.537567  &  49.254562  &  0.037  &      9.46  &    -1.28  &    2.4  &       2.3  &  41.0  &       AGN  &           AGN  \\
        8992-3702  &  171.657262  &  51.573041  &  0.026  &      9.47  &    -1.49  &    1.2  &       2.4  &  40.2  &       AGN  &           AGN  \\
  9000-1901$^{*}$  &  171.400654  &  54.382574  &  0.021  &      9.27  &    -0.65  &   -2.5  &       1.0  &  40.2  &     LINER  &           AGN  \\
        9031-1902  &  241.029075  &  44.549765  &  0.043  &      9.45  &    -1.27  &   -2.0  &       1.6  &  40.9  &       AGN  &           AGN  \\
        9033-3702  &  222.337719  &  47.417495  &  0.026  &      9.37  &    -1.54  &   -0.7  &       2.2  &  39.5  &     LINER  &     Composite  \\
        9035-1901  &  235.405343  &  44.272184  &  0.037  &      9.14  &    -2.32  &   -1.5  &       1.4  &  39.7  &     LINER  &     Quiescent  \\
        9488-1901  &  126.409848  &  21.142559  &  0.023  &      9.26  &    -2.19  &   -2.0  &       1.2  &  40.1  &     LINER  &           AGN  \\
        9865-1902  &  223.866232  &  51.047738  &  0.030  &      9.13  &    -2.50  &   -0.6  &       1.8  &  40.1  &       AGN  &           Quiescent  \\
        9881-1901  &  204.806913  &  24.893076  &  0.028  &      9.29  &    -1.88  &    0.4  &       1.6  &  40.4  &       AGN  &             -  \\
\hline
\hline
\end{tabular}
\smallskip\newline\small {\bf Column designation:}~(1) MaNGA plateifu; (2,3) RA, DEC coordinates of the optical center of the galaxy or IFU center; (4) galaxy redshift; (5) galaxy stellar mass; (6) galaxy star-formation rate within the IFU field-of-view based on the H$\alpha$ flux; (7) galaxy TType ($>$0 for late-type, $<0$ for early-type); galaxy effective radius; (8) bolometric luminosity of the AGN/LINER spaxels (derived from their median [OIII] luminosity); (9) MaNGA (IFU) median BPT classification (derived from the median fluxes of the AGN/LINER spaxels); (9) SDSS (single-fiber) BPT classification. $^{*}$Galaxy with a FIRST radio counterpart.
\end{table*}

\subsection{Host galaxy properties}
\label{host}
The photometric properties of the dwarf galaxies are drawn from the MaNGA PyMorph  photometric value added catalog (MPP-VAC; \citealt{2019MNRAS.483.2057F}), who performed Sersic and Sersic+Exponential fits to the 2D surface brightness profiles of the MaNGA DR15 galaxy sample. The morphological properties are drawn from the companion catalog, the MaNGA deep learning morphological catalog (MDLM-VAC; \citealt{2019MNRAS.483.2057F}), which provides a morphological classification of the same galaxy sample based on the deep-learning algorithm presented in detail in \cite{2018MNRAS.476.3661D}. The surface brightness profiles are available for the 37 dwarf galaxies with AGN candidates, 29 of which are best fitted with a Sersic+Exponential profile or with a Sersic profile with index $\leq$2. This is in agreement with the late-type classification (i.e. TType $>$ 0 in the MDLM-VAC catalog) of 25 of these sources by the deep-learning based algorithm. The remaining dwarf galaxies are classified as early-type (TType $<$ 0). 
The B-V colors derived from the surface brightness profiles range from 0.7 to 1.1, with a median value $<B-V>$ = 0.9. 
Such colors are redder than those found by single-fiber optical searches of AGN in dwarf galaxies (e.g. \citealt{2013ApJ...775..116R}) and also those of X-ray and radio searches (e.g. \citealt{2018MNRAS.478.2576M,2019MNRAS.488..685M}), and are in agreement with the finding that only six out of the 37 IFU AGN candidates are classified as star-forming according to the integrated SDSS BPT (see Table~\ref{table1}). A predominance of star-formation over AGN ionisation in the central regions seems thus not to be the cause of single-fiber optical diagnostics not being able to fully recover AGN emission in 23 of the 37 AGN candidates. Instead, the AGN emission in most of these sources must be either off-nuclear or turned-off.

\subsection{Other wavelengths}
\label{radio}
To distinguish between star-formation dilution of the AGN ionisation, turned-off AGN, and true off-nuclear wanderers we search for AGN signatures in the mid-infrared, radio, and X-rays. The 37 dwarf galaxies have mid-infrared WISE counterparts; however only two of them are classified as an AGN according to the color cuts of \cite{2011ApJ...735..112J}: 9000-1901 and 8982-3703, already classified as AGN based on the SDSS spectrum. In the X-ray regime, no \textit{Chandra} data is available.

In the radio, seven of the 37 dwarf galaxies have a 1.4 GHz counterpart in the FIRST\footnote{Faint Images of the Radio Sky at Twenty Centimeters.} survey (see Table~\ref{table1}). The peak radio emission ranges from $\sim$1.1 to 20.4 mJy beam$^{-1}$ and is in all cases except for one (8442-1901) resolved. We derive the expected thermal and non-thermal contribution from star formation to the detected 1.4 GHz FIRST radio emission using the correlation for dwarf galaxies of \citeauthor{2019MNRAS.484..543F} (2019; see also \citealt{2019MNRAS.488..685M}). We find that the FIRST radio emission is $\geq$ 3$\sigma$ larger than that expected from star formation for all the sources, whose star formation contribution to the radio emission ranges from 0.2\% to 20.8\%. At least $\sim$80\% of the radio flux is thus expected to come from an AGN. Most of the dwarf galaxies with FIRST radio emission lack AGN/LINER spaxels at the center of the galaxy, which suggests that either the AGN ionisation is diluted by star formation in the central region or that the AGN is switched off. The later can be now strongly disfavored based on the finding of central AGN radio emission. 
We note that four out of the seven sources with radio emission are classified as LINER in the spaxel-averaged BPT, which reinforces not having excluded LINERs from the final sample of AGN dwarf galaxy candidates (Sect.~\ref{sample}). Higher-resolution radio observations are planned for the sources with a FIRST counterpart in order to confirm the presence of an AGN.

\subsection{Off-nuclear wanderers, hidden, or switched-off AGN?}
While the AGN/LINER spaxels of the 14 sources classified as AGN by the SDSS data are centrally located on the IFU center, for the 23 new AGN candidates most of the AGN/LINER spaxels are offset from the optical center. In the cases with clearly off-nuclear AGN-like line ratios, the emission tends to be diffuse, elongated and often symmetric, consistent with being light echoes of a past AGN activity (e.g. 8083-3704, 8442-1901, 8465-9101; see Fig.~\ref{fig1}). The switched-off AGN scenario is however very unlikely for those sources whose FIRST radio emission is consistent with an AGN origin. In these cases the off-nuclear location of the AGN/LINER spaxels can be rather explained by either star-formation dilution (e.g. 8255-12704, 8442-1901, 8465-9101, 8932-9102, 8932-9102; see Fig.~\ref{fig1}), as supported by the star-forming classification of some of these sources, or the AGN being off-nuclear (e.g. 8932-9102; see Fig.~\ref{fig1}). 

For the off-nuclear AGN candidate 8932-9102, the location of the AGN with respect to the optical center can be derived as the median of the distance between each AGN/LINER spaxel and the IFU center. We find an offset of 6.4 arcsec, which corresponds to 2.7 kpc. This offset is consistent with those found in radio (\citealt{2019MNRAS.488..685M}; \citealt{2020ApJ...888...36R}) and X-ray searches (e.g. \citealt{2018MNRAS.478.2576M}) of AGN in dwarf galaxies, and with the (low-luminosity) AGN nature of some ultraluminous X-ray sources (e.g. \citealt{2017ApJ...844L..21K}; \citealt{2013MNRAS.436.1546M,2013MNRAS.436.3128M,2015MNRAS.448.1893M,2018MNRAS.480L..74M}; \citealt{2019ApJ...882..181B}). The offset is also in agreement with simulations of seed BH formation, which predict that a population of wandering BHs should be present in dwarf galaxies if these have undergone interactions and mergers (e.g. \citealt{2019MNRAS.482.2913B}). 

\subsection{AGN luminosity}
\begin{figure}
\centering
\includegraphics[width=0.48\textwidth]{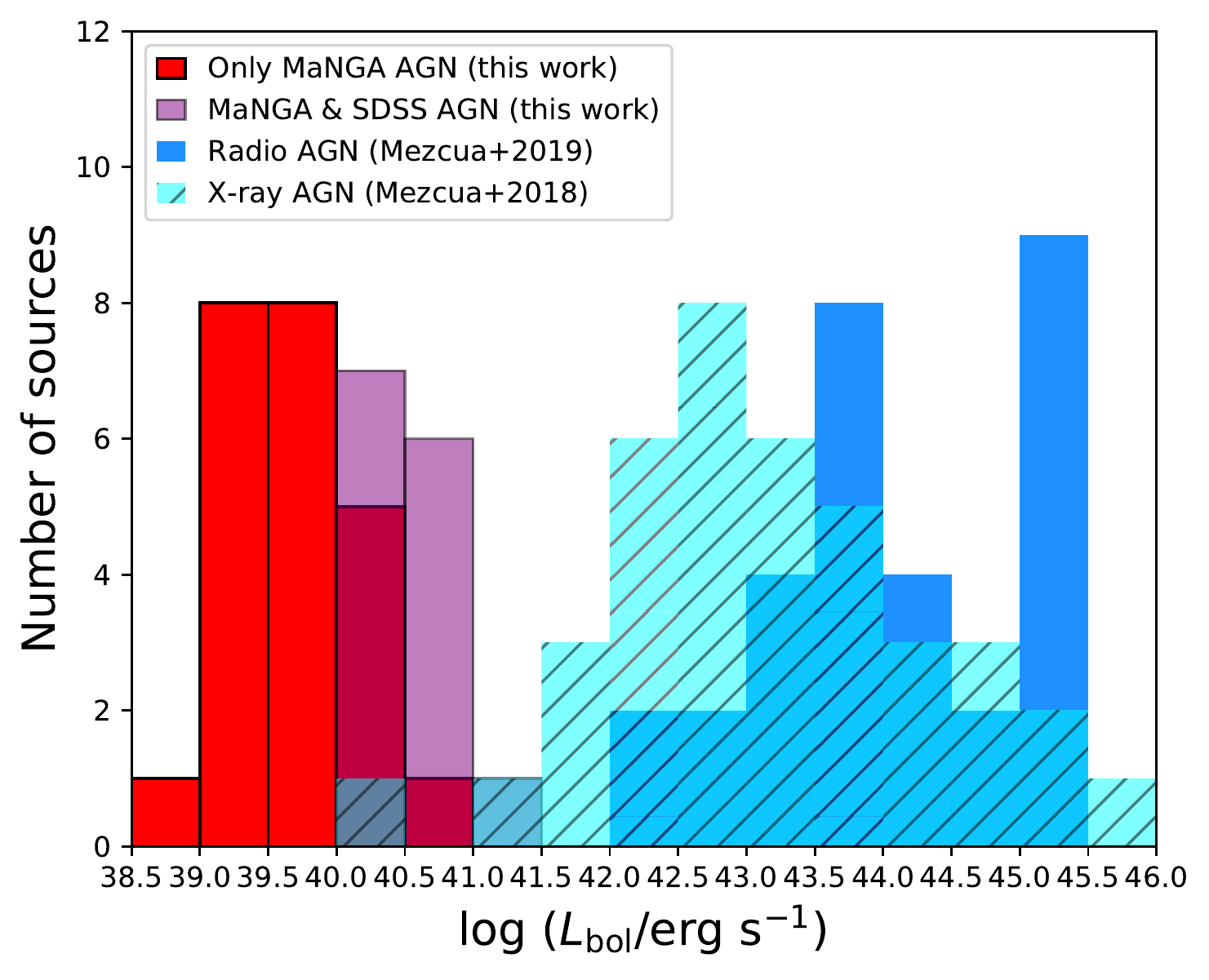}
\caption{Distribution of bolometric luminosity for the 23 new MaNGA AGN candidates (red bars), the 14 SDSS AGN (purple bars), the sample of radio AGN dwarf galaxies from \citeauthor{2019MNRAS.488..685M} (2019; blue solid bars) and the X-ray sample of AGN dwarf galaxies from \citeauthor{2018MNRAS.478.2576M} (2018b; cyan hashed bars).}
\label{Lbol}
\end{figure}

The bolometric luminosity ($L_\mathrm{bol}$) of the 37 AGN candidates can be estimated from the [OIII] luminosity ($L_\mathrm{[OIII]}$). The later is derived from the median [OIII] flux of the AGN/LINER spaxels and converted to $L_\mathrm{bol}$ assuming a bolometric correction of 1000 (\citealt{2014AJ....148..136M}). We find a range of log $L_\mathrm{bol}$ =  38.9-41.4 erg s$^{-1}$, which is below the typical AGN bolometric luminosity threshold of $\sim10^{42}$ erg s$^{-1}$ and indicates that these are low-luminosity AGN (e.g. \citealt{2014ApJ...787...62M}). We note that excluding those sources classified as AGN by the SDSS BPT the median bolometric luminosity would be of log $L_\mathrm{bol}$ =  39.6 erg s$^{-1}$ (see Fig.~\ref{Lbol}), which is nearly one order of magnitude lower than the median log $L_\mathrm{bol}$ =  40.5 erg s$^{-1}$ of the 14 SDSS AGN. The $L_\mathrm{bol}$ is also more than two orders of magnitude lower than those of X-ray and radio-selected AGN (Fig.~\ref{Lbol}; e.g. \citealt{2018MNRAS.478.2576M,2019MNRAS.488..685M}), which highlights the power of IFU spectroscopy in revealing very faint AGN in dwarf galaxies.

\subsection{Black hole mass: \\revealing a hidden type 1 low-mass AGN}
For three of the SDSS AGN a BH mass measurement is reported in the literature (e.g. \citealt{2015ApJ...813...82R}) based on the width of broad H$\alpha$ and H$\beta$ emission lines using standard virial techniques: 8446-1901 (log $M_\mathrm{BH}$ = 6.52 M$_{\odot}$), 8992-3702 (log $M_\mathrm{BH}$ = 5.99 M$_{\odot}$), and 9000-1901 (log $M_\mathrm{BH}$ = 7.27 M$_{\odot}$). 

The dwarf galaxy 8442-1901 also presents broad H$\alpha$ emission in the stacked AGN MaNGA spectrum (see Figure \ref{8442-1901}), however such broad component has not been accounted for in the literature (most likely because of its Star-forming/Composite BPT classification based on the single-fiber SDSS spectrum) nor is it recovered by the MaNGA data-analysis pipeline (see Appendix \ref{broadlines}). To confirm the AGN BPT classification of 8442-1901 by the MaNGA data we thus perform our own emission line fitting of the H$\alpha$ + [NII] complex (see description in Appendix \ref{broadlines}). As a result we find a broad H$\alpha$ emission line component of FWHM = 1131 $\pm$ 378 km s$^{-1}$, from which we derive a BH mass log $M_\mathrm{BH}$ = 5.07 M$_{\odot}$ (with a typical uncertainty of 0.5 dex) in the IMBH regime. The AGN nature of 8442-1901 is further reinforced by the finding that only 0.9\% of its compact FIRST radio emission originates from star formation (see Sect.~\ref{radio}). This indicates that the symmetric, elongated off-nuclear morphology of the AGN spaxels are not light echoes from a switched-off AGN. Instead, the MaNGA data reveal a hidden type 1 AGN in a dwarf galaxy whose central AGN emission is outshined by star formation in the single-fiber (SDSS) BPT diagnostic diagram. 

For the remaining 33 AGN dwarf galaxy candidates with no broad H$\alpha$ or H$\beta$ emission lines we estimate the BH mass from the stellar velocity dispersion ($\sigma$) within 1 $R_\mathrm{eff}$ applying the $M_\mathrm{BH}-\sigma$ correlation (e.g. \citealt{2018ApJ...855L..20M}; \citealt{2019arXiv191109678G}), which should be less biased than the $M_\mathrm{BH}-M_\mathrm{*}$ correlation (e.g. \citealt{2019MNRAS.485.1278S}; \citealt{2020MNRAS.491.1311M}). The low-mass end of the $M_\mathrm{BH}-\sigma$ correlation is found to flatten in some studies (\citealt{2006ApJ...641L..21G}; \citealt{2017IJMPD..2630021M}; \citealt{2018ApJ...855L..20M}) but to remain unchanged in others (e.g. \citealt{2019arXiv191109678G}), which can be explained by a possible bias towards massive BHs when using only detections (i.e. excluding upper limits) on BH mass measurements (\citealt{2019arXiv191109678G}). We thus estimate the BH mass using both possibilities: using the correlation from \cite{2018ApJ...855L..20M} we find a range 
log $M_\mathrm{BH}$ = 5.6-6.6 M$_{\odot}$, with 14 sources qualifying as IMBHs (log $M_\mathrm{BH} < 6$ M$_{\odot}$); while using \cite{2019arXiv191109678G} we find log $M_\mathrm{BH}$ = 5.8-7.9 M$_{\odot}$ with three sources qualifying as IMBHs.
In both cases the Eddington ratios derived as $\lambda_\mathrm{Edd} = L_\mathrm{bol}/(M_\mathrm{BH} \times 1.3 \times 10^{38})$ are $\lambda_\mathrm{Edd} \leq$ 1\%, indicating that all the sources are accreting at sub-Eddington rates.

\subsection{Demographics}
\label{demographics}
Using MaNGA IFU we find 37 AGN candidates among a sample of 1609 dwarf galaxies with $M_\mathrm{*} < 3 \times 10^{9}$ M$_{\odot}$, which translates into an AGN fraction of 2.3\%. This fraction is smaller than the 5.1\% of \cite{2018MNRAS.474.1499W}, who investigated 2727 galaxies observed by MaNGA (half of our initial 4718 sources) and found 48 AGN in dwarf galaxies with $M_\mathrm{*} < 3 \times 10^{9}$ M$_{\odot}$ (Petrosian NSA masses using h=0.73). Although \cite{2018MNRAS.474.1499W} perform a stringent AGN identification based also on the [NII]-BPT, [SII]-BPT, and WHAN diagrams, many of their AGN candidates are not included in our sample due to their low number of AGN/LINER spaxels. \cite{2017MNRAS.472.4382R} and \cite{2018RMxAA..54..217S} also searched for AGN using MaNGA, but both works focus on the central region of galaxies and thus do not include most of the light-echoes and off-nuclear AGN dwarf galaxy candidates found here. \cite{2018MNRAS.476..979P} used the MaNGA survey to investigate AGN in dwarf galaxies, but they performed a different procedure than us, applying first the WHAN diagram to the whole integrated spectrum covered by MaNGA (and not only to the AGN/LINER spaxels) in order to identify quenched galaxies and plotting later the BPT diagram in order to find AGN in these sources. As a result, only one (8982-3703) of their six AGN dwarf galaxies is included in our final sample of AGN candidates. 

Among single-fiber spectroscopic studies, one of the largest searches for AGN in dwarf galaxies is that of \cite{2013ApJ...775..116R}, who identified 136 AGN/composite objects in the [NII]- and [SII]-BPT diagrams among a sample of 25974 SDSS dwarf galaxies with $M_\mathrm{*} < 3 \times 10^{9}$ M$_{\odot}$. Excluding composite objects, only 17 of their AGN qualify as such by the WHAN diagram, which translates into an AGN fraction of $\sim$0.1\%. This is significantly smaller than our AGN fraction, which we attribute to the bias of single-fiber spectroscopy towards central AGN. Indeed, considering only the 14 SDSS AGN, we would have an AGN fraction of $\sim$0.9\%, closer to that of the single-fiber SDSS studies but lower than that of the whole MaNGA AGN dwarf galaxy sample, which emphasizes the power of IFU in demographic studies of AGN in dwarf galaxies.

\section{Conclusions} 
\label{conclusions}
Using MaNGA IFU data we have identified a sample of 37 AGN candidates in dwarf galaxies. In 23 of them the AGN ionisation signatures are largely missed by single-fibre (SDSS) spectroscopy, which classifies them as either Quiescent, Star-Forming or Composite. While this suggests that star formation is the dominant source of ionisation in the central region, most of the galaxies are not star-forming. The AGN could thus be either switched off or wandering in the host dwarf galaxy, as predicted by models of seed BH formation. A BH mass $\sim10^5$ M$_{\odot}$ is derived for one of the 23 new AGN candidates based on the detection a broad H$\alpha$ emission line component, which constitutes the first hidden type 1 low-mass AGN to be revealed by IFU data.
The AGN bolometric luminosity of the 23 new AGN candidates is $\lesssim 10^{40}$ erg s$^{-1}$, nearly one order of magnitude fainter than that of the dwarf galaxies with single-fiber AGN emission and several orders of magnitude lower than that of X-ray and radio AGN dwarf galaxies. IFU surveys offer thus a potential tool for identifying hidden and faint AGN, which is crucial for population studies of AGN in dwarf galaxies and for understanding whether these host the seed BHs of the early Universe. The detailed kinematic an stellar population properties of the sample will be presented in a fore-coming paper, while radio and X-ray observations are planned to confirm the AGN nature.

\acknowledgments
The authors would like to thank they anonymous referee for his/her helpful comments. MM would like to thank A. Olmo-Garc\'ia for sharing her emission line fitting code, S. S\'anchez for helpful discussions, and D. Wylezalek for sharing the stellar masses of her AGN sample. HDS thanks M. Bernardi for her help with the stacking spectra procedure. MM acknowledges support from the Beatriu de Pinos fellowship (2017-BP-00114). HDS acknowledges support from the Centro Superior de Investigaciones Cient\'ificas PIE2018-50E099.



\bibliographystyle{aasjournal}
\bibliography{/Users/mmezcua/Documents/referencesALL}

\clearpage
\appendix

\section{Stacking}
\label{stacking}
 For each individual galaxy, the spectra of all the spaxels which fall in the AGN/LINER region are stacked together. We use the HYB10-GAU-MILESHC maps from the MaNGA DR15, which provide the observed flux, the emission line component, the stellar continuum and the best-fitting model. We stack both the observed flux (in blue in figures \ref{fig1},  \ref{8442-1901} and  \ref{8446-1901}) and the emission line components (shown in red) to better visualise the emission line fluxes. The stacking procedure is performed as follows: the spectra are first rest-frame corrected (including the redshift and the correction due to the rotational velocity) and then the wavelengths are converted to air units following the standard convention. The flux is normalised in two regions free of emission lines in the blue and red part of the spectrum (4600 - 4700 \AA ; 6780 - 6867 \AA) to remove the intrinsic slope of the spectra. The stacked spectrum is computed as the median value of the $3\sigma$ clipped flux at each wavelength. The emission line component is shifted by 1 to match the observed flux. Skylines or masked wavelength regions are not used in the stacking (see \citealt{2019MNRAS.489.5612D} for additional details). 

\section{Broad emission line measurements}
\label{broadlines}
\begin{figure*}
\centering
\includegraphics[width=\textwidth]{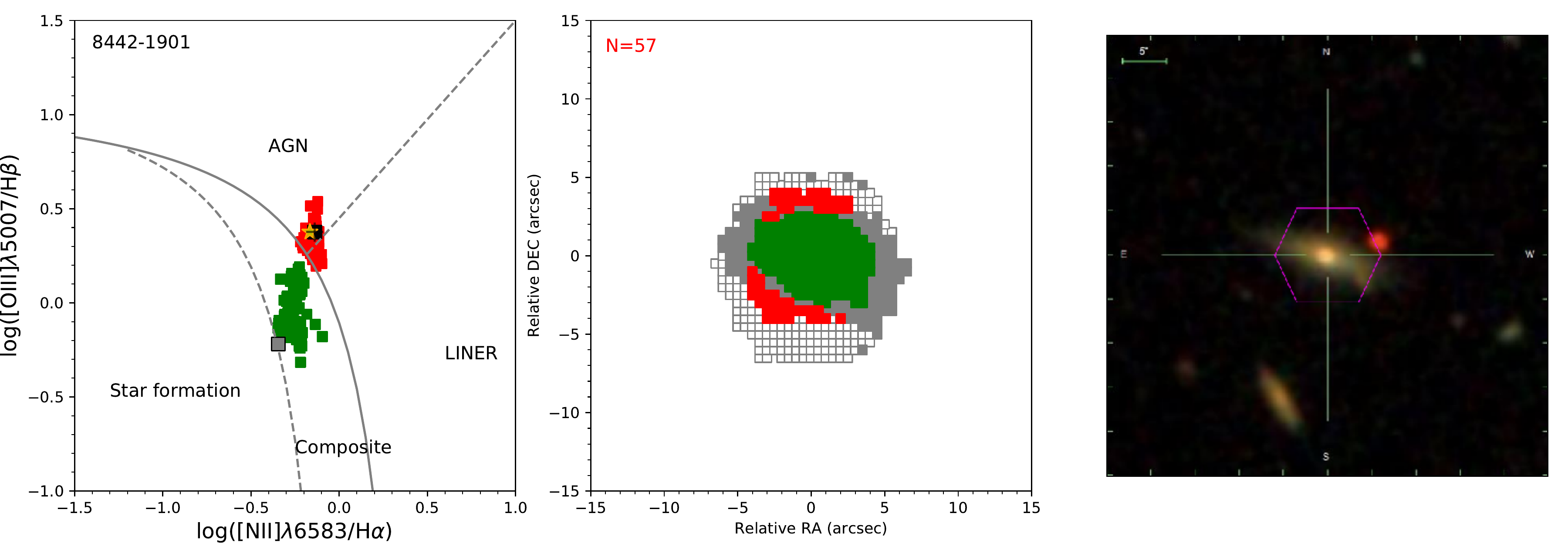}
\includegraphics[width=\textwidth]{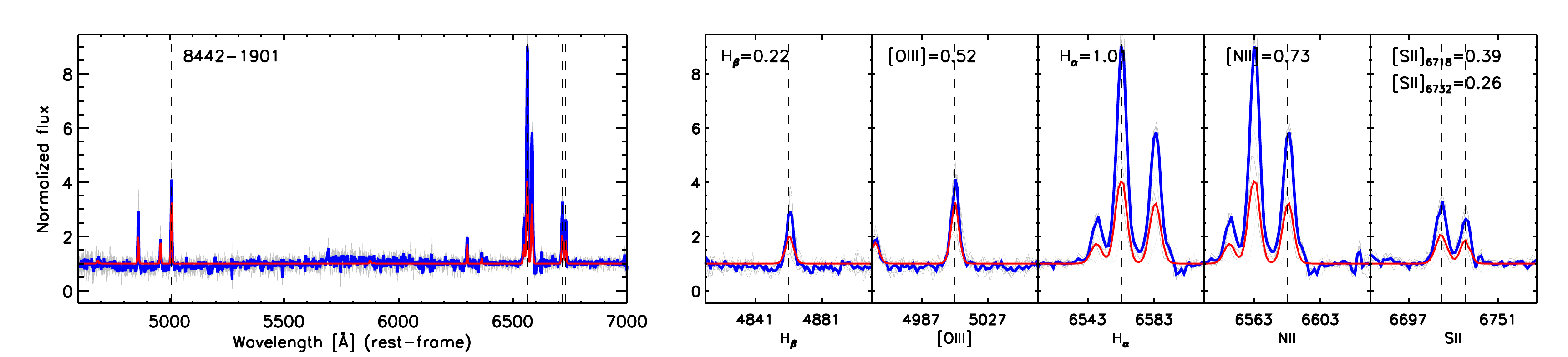}
\includegraphics[width=0.5\textwidth]{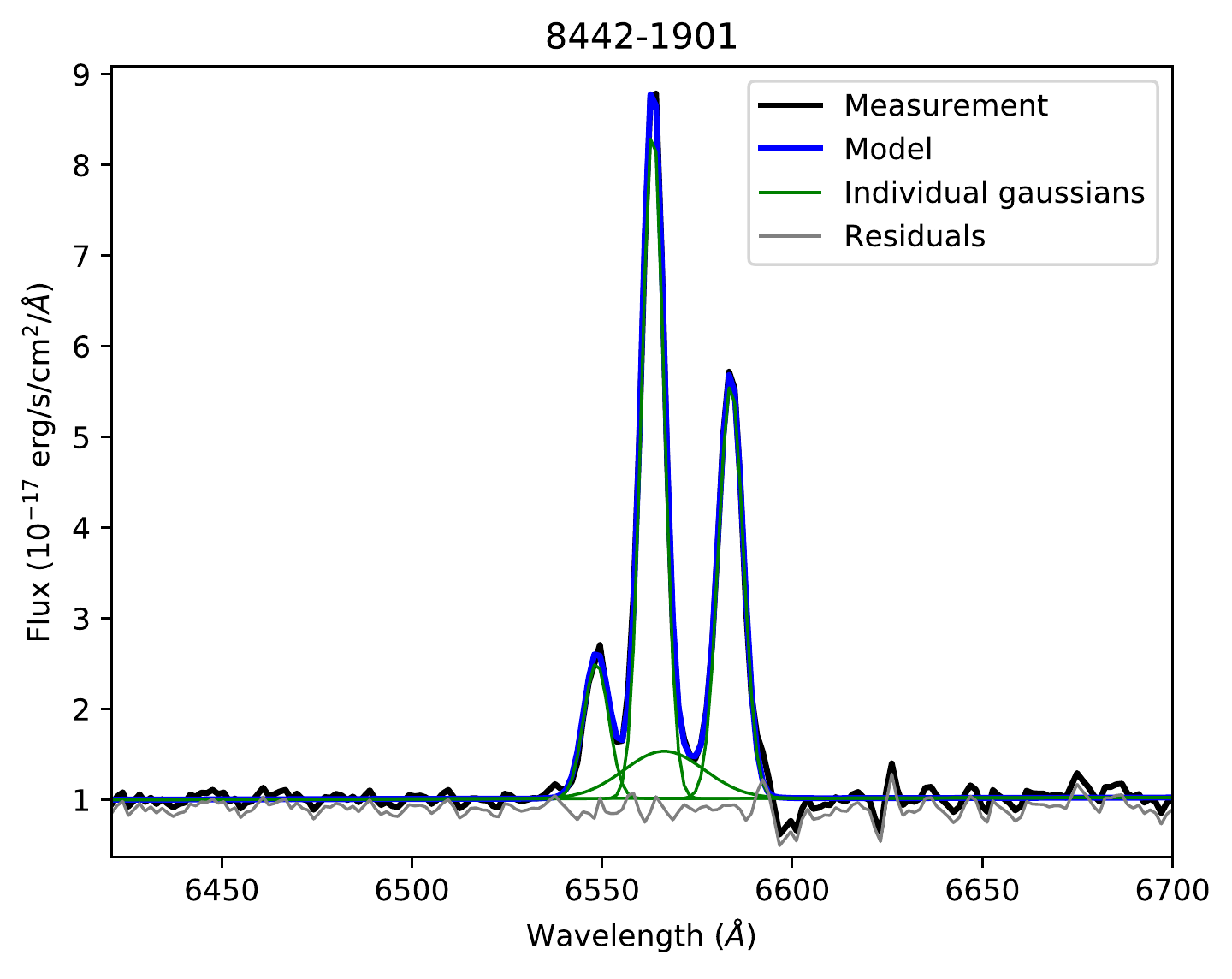}
\caption{Same caption as in Fig.~\ref{fig1}. The yellow star marks the median BPT location based on the fluxes derived from the multi-Gaussian fitting of the stacked spectrum of the AGN/LINER spaxels. \textbf{Bottom}: Multi-Gaussian emission line fitting of the H$\alpha$+[NII] complex of the stacked spectrum.}
\label{8442-1901}
\end{figure*}

We perform a visual inspection of all the AGN/LINER candidates in order to check the consistency of the emission line measurements provided by MaNGA data-analysis pipeline. The visual inspection of the stacked spectrum of 8442-1901 shows that the MaNGA emission line model seems not to correctly fit the emission lines used in the BPT diagnostic diagram, which could be biasing its BPT classification as an AGN (see Fig.~\ref{8442-1901}). This is likely due to the possible presence of a broad H$\alpha$ emission line component in the stacked AGN spectrum of 8442-1901. To probe this, we perform a multi-Gaussian fit to the H$\alpha$+[NII] complex (and also of the H$\beta$ and [OIII] emission lines used in the BPT diagram) using the Python package LMFIT as in \cite{2017ApJ...834..181O}. We follow the common procedure (e.g. \citealt{2013ApJ...775..116R}) of assuming the same width for the [NII]$\lambda$6583, [NII]$\lambda$6548, [OIII]$\lambda$5007, [OIII]$\lambda$4958, narrow H$\alpha$, and narrow H$\beta$ emission line Gaussian components and allowing for a secondary broad H$\alpha$ and H$\beta$ component. The relative separations between the centers of the narrow components are fixed using their theoretical wavelengths and the flux ratios of the [NII] and [OIII] doublets are set to the theoretical values of 3.06 and 2.99, respectively. The best model (that with the lowest $\chi$$^{2}$) is obtained when including a broad H$\alpha$ component but no H$\beta$ broad component (see Fig.~\ref{8442-1901}, bottom). The resulting FWHM is corrected for instrumental resolution, thermal and natural broadening, which yields a FWHM (broad H$\alpha$) = 1131 $\pm$ 378 km s$^{-1}$. This, together with the luminosity (log L(H$\alpha$) = 38.6 erg s$^{-1}$) derived from the flux of the broad H$\alpha$ component, allows us to derive a BH mass under the assumption that the broad line region from which the broad H$\alpha$ originates is virialized. Using a geometrical factor $\epsilon$ = 1 (eq. 5 in \citealt{2013ApJ...775..116R}), we find log $M_\mathrm{BH}$ = 5.07 M$_{\odot}$ (with a typical uncertainty of 0.5 dex). 
The resulting BPT classification, based on the derived narrow emission lines fluxes of the H$\alpha$, H$\beta$, [NII]$\lambda$6583, and [OIII]$\lambda$5007, is shown as a yellow star in Fig.~\ref{8442-1901} (top left panel) and confirms the AGN classification of 8442-1901.

\begin{figure*}
\centering
\includegraphics[width=\textwidth]{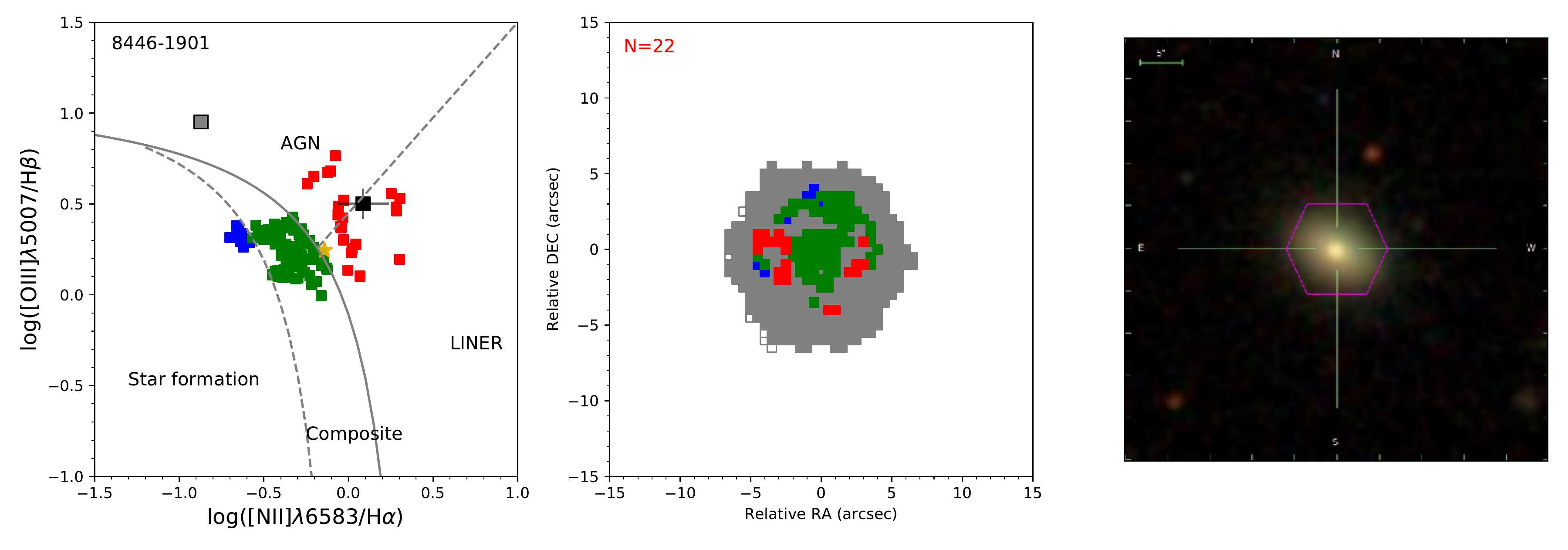}
\includegraphics[width=\textwidth]{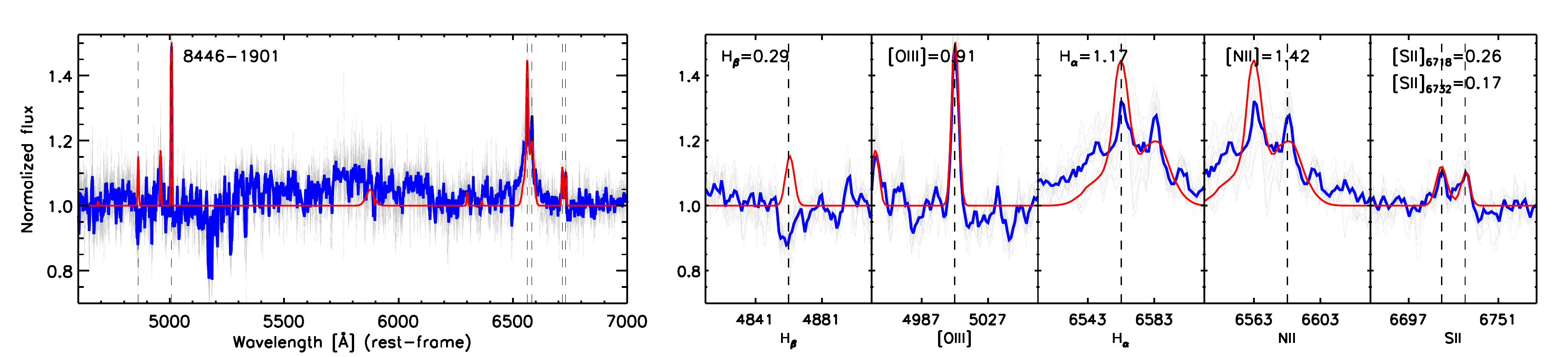}
\includegraphics[width=0.5\textwidth]{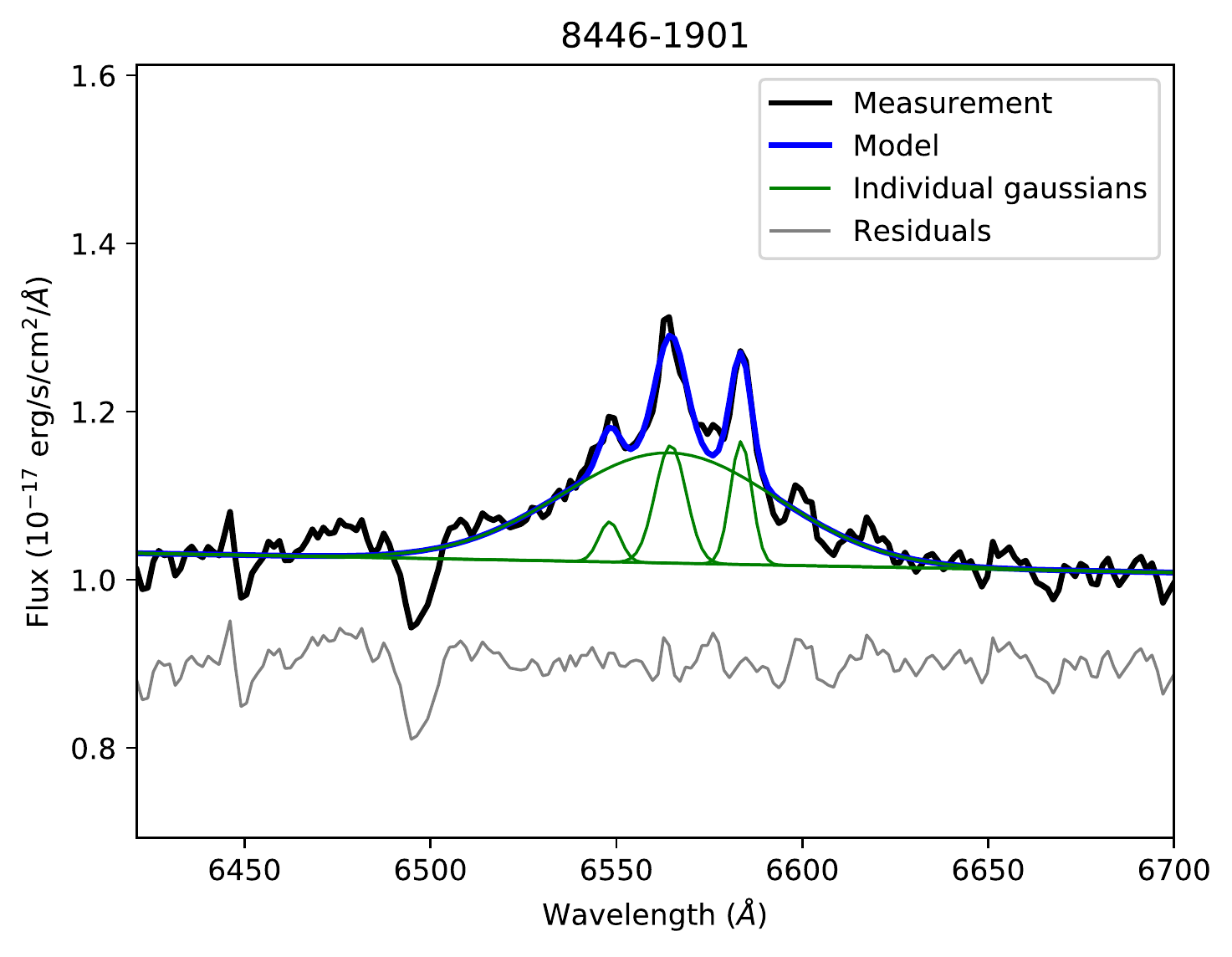}
\caption{Same caption as in Fig.~\ref{fig1}. The yellow star marks the median BPT location based on the fluxes derived from the multi-Gaussian fitting of the stacked spectrum of the AGN/LINER spaxels. \textbf{Bottom}: Multi-Gaussian emission line fitting of the H$\alpha$+[NII] complex of the stacked spectrum.}
\label{8446-1901}
\end{figure*}

For 8446-1901 the MaNGA emission line model seems, as for 8442-1901, not to correctly fit the H$\alpha$+[NII] emission line complex due to the presence of a broad H$\alpha$ emission line component already present in the SDSS spectrum (see Fig.~\ref{8446-1901}). To confirm the AGN BPT classification of MaNGA we thus also perform a multi-Gaussian fit of the H$\alpha$+[NII] complex that includes a broad H$\alpha$ component (Fig.~\ref{8446-1901}, bottom). We find FWHM (broad H$\alpha$) = 3116 $\pm$ 205 km s$^{-1}$ and log $M_\mathrm{BH}$ = 5.73 M$_{\odot}$. Using the narrow emission line fluxes of H$\alpha$ and [NII]$\lambda$6583, 8442-1901 is classified as an AGN/LINER in the BPT diagram (yellow star in the top left panel of Fig.~\ref{8446-1901}).

The BPT diagram and stacked spectrum of the 31 AGN dwarf galaxy candidates not shown here nor in the main paper will be made available in the online version of the manuscript.

\end{document}